\documentclass[epsf,prb,twocolumn,showpacs,preprintnumbers]{revtex4}
\usepackage{graphics}
\usepackage{graphicx}
\usepackage{dcolumn} 
\usepackage{bm}
\usepackage{color}
\usepackage{epsfig}
\begin{document}
\title{ Half-metallic Ferrimagnetism Driven by Coulomb Enhanced 
Spin-Orbit Coupling in PdCrO$_3$
}
\author{Hyo-Sun Jin$^1$ and Kwan-Woo Lee$^{1,2}$} 
\affiliation{ 
$^1$Department of Applied Physics, Graduate School, Korea University, Jochiwon, 
  Chungnam 339-700, Korea \\
 $^2$Department of Display and Semiconductor Physics, Korea University, Jochiwon, 
  Chungnam 339-700, Korea
}
\date{\today}
\pacs{71.20.Be, 71.27.+a, 75.47.Np}
\begin{abstract}
Recently, in the seemingly narrow gap insulating NiCrO$_3$ 
with the trigonally distorted ($R\bar{3}c$) perovskite-like structure,
a compensated half-metal (CHM) is predicted, as applying a modest pressure. 
Using {\it ab initio} calculations including both Coulomb correlations 
and spin-orbit coupling (SOC), we investigate the as-yet-unsynthesized PdCrO$_3$,
isostructural and isovalent to NiCrO$_3$.
Upon applying the on-site Coulomb repulsion $U$ to both Pd and Cr ions,
the Cr spin moment is precisely compensated with the antialigned spin moments
of Pd and oxygens. Coincidentally only one spin channel remains metallic due to
the twice larger width of the Pd $4d$ bands than the Ni $3d$ bands in NiCrO$_3$, 
indicating CHM in ambient pressure. 
Inclusion of SOC as well as correlation effects (LDA+U+SOC) produces 
a SOC constant enhanced twice over the value of LDA+SOC, 
leading to unusually large orbital moment of --0.25 $\mu_B$ on Pd.
However, the half-metallicity still survives, 
so that a transition of CHM to a half-metallic ferrimagnet occurs
due to Coulomb enhanced SOC.
On the other hand, an isovalent, but presumed cubic double perovskite
La$_2$PdCrO$_6$ is expected to be a half-metal ferromagnet 
with tiny orbital moments.
\end{abstract}
\maketitle

\begin{figure}[tbp]
{\resizebox{5.5cm}{8cm}{\includegraphics{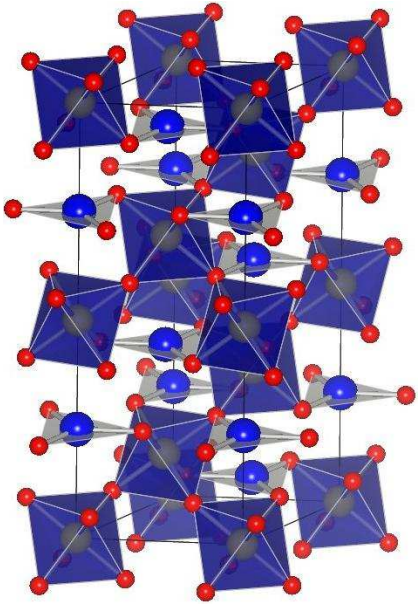}}}
\caption{(Color online) Our optimized crystal structure ($R\bar{3}c$) 
of the trigonally distorted PdCrO$_3$, 
consisting of CrO$_6$ octahedra and PdO$_3$ triangular
planes: the small (red) spheres  for O and the large
(blue) spheres for Pd. 
The PdO$_3$ triangular planar symmetry leads to the crystal field
splittings: doublets $e_{1g}$ ($d_{yz}$, $d_{zx}$), $e_{2g}$ ($d_{xy}$, $d_{x^2-y^2}$), 
and singlet $a_{1g}$ ($d_{z^2}$).
This figure was produced by VESTA.\cite{vesta}
}
\label{str}
\end{figure}

A half-metal shows spin-dependent transport phenomena, ideally 100\%
spin polarization defined as 
\begin{eqnarray}
 P=\frac{N_\uparrow(E_F)-N_\downarrow(E_F)}{N_\uparrow(E_F)+N_\downarrow(E_F)},\nonumber
\end{eqnarray}
where $N_\uparrow(E_F)$ and $N_\downarrow(E_F)$ are the spin up
and down densities of states at the Fermi energy $E_F$, respectively.
This is extremely interesting as a promising candidate of a spin injector 
in spintronics.\cite{groot0}
In the early 1980's, de Groot and coworkers suggested a half-metallic 
ferromagnet (HMFM).\cite{groot1} 
Focusing on stoichiometric compounds, HM has an integral net moment 
due to the nature of HM.
Additionally, half-metallic ferrimagnets (HMFI) have been searched,
since a higher Curie temperature has been expected.\cite{groot2,chioncel,lv}
However, application of these kinds of HMs to spintronics
has confronted some problems, in particular due to the presence of stray-field.
Thus, for about the last 20 years, HMs with zero net magnetic moment,
compensated half-metals (also so-called half-metallic 
antiferromagnets),\cite{felser} have been sought.
Since a compensated half-metal (CHM) was suggested 
in a Heusler compound,\cite{groot3}
double perovskites,\cite{wep1,park,LP1,PP,song,chen} 
tetrahedrally coordinated chalcopyrites,\cite{nakao,long}
and iron-pnictides\cite{hu} have been theoretically studied.
In spite of a number of theoretical 
prediction,\cite{felser,groot3,wep1,park,LP1,PP,song,chen,nakao,long,hu,erosy,LP2} 
no true CHM has been synthesized yet.

Very recently, Lee and Pickett suggest
that applying moderate pressure leads to CHM in trigonally
distorted perovskite-type NiCrO$_3$,
which seems to be a semiconductor with the very small gap 
in the spin up channel in ambient pressure.\cite{LP2}
In this paper, we will investigate an isostructural and isovalent
PdCrO$_3$, but unsynthesized yet, using first principles calculations
including both Coulomb correlations and spin-orbit coupling (SOC).
Since strength of correlation of Pd ions is much weaker than that of Ni ions,
a (half) metallic PdCrO$_3$ is expected even in ambient pressure.
Although a Pd ion is often nonmagnetic, 
a spin ordered Pd ion has not been rarely observed.\cite{hong,kim,rod}

Compared with $5d$ or $f$ systems, $3d$ and $4d$ systems show relatively
small SOC. Combined with Coulomb correlations, however, SOC
is sometimes amplified, so called ``Coulomb enhanced SOC".\cite{srrho} 
In the spinel compounds $\cal{T}$V$_2$O$_4$ ($\cal{T}$=Zn, Mn),
Coulomb enhanced SOC leads to an orbital ordering on V ions
with large orbital moments of 0.34 or 0.75 $\mu_B$.\cite{znvo,mnvo}
In another spinel FeCr$_2$S$_4$, a considerable orbital moment of 0.14 $\mu_B$ 
on Fe ions was obtained in LDA+U+SOC calculations.\cite{fecrs}

Usually, strong SOC induces mixing of states to the opposite spin channel,
resulting in undermining half-metallicity.\cite{LP1,wep2,ph}
On the contrary, in the case of PdCrO$_3$ showing Coulomb
enhanced SOC, a considerable orbital moment is induced, but
the mixing of states is tiny.
Considering only correlation effects, PdCrO$_3$ is a precise CHM.
Our LDA+U+SOC calculations produce substantial
orbital moment of $-0.25$ $\mu_B$ on Pd ion,
resulting in nonzero net moment ($-0.29$ $\mu_B$/f.u.).
However, the half-metallicity remains unchanged.
As a result, this system becomes a half-metallic ferrimagnet. 
HF with nonintegral net moment is unprecedented for stoichiometric compounds, 
as far as we know.
On the other hand, our calculations suggest that the orbital moments are tiny
in a presumed cubic double perovskite La$_2$PdCrO$_6$, 
isovalent to PdCrO$_3$, so being HMFM.

We fully optimized both the cubic and the trigonally distorted
perovskite ($R\bar{3}c$) structures.
For the distorted structure, however, the ratio of lattice parameters
$\frac{a}{c}$ was kept to be the same as in NiCrO$_3$.\cite{LP2}
As observed in NiCrO$_3$,\cite{LP2}
our relaxations show that the distorted structure ($R\bar{3}c$) 
is energetically favored by 0.3 eV per formula unit (f.u.)
over the cubic structure. 
The lattice parameters in the distorted phase are
$a$=4.9597~\AA~ and $c$=13.5985~\AA, close to the volume of the cubic phase.
Thus, we will address only the distorted phase.
In the $R\bar{3}c$ structure, O ions lie on $18e$ ($x$, 0, $\frac{1}{4}$)
sites with the optimized internal parameter $x$=0.5845.
Pd and Cr ions sit on the $6a$ (0,0,$\frac{1}{4}$) and $6b$ (0,0,0),
respectively.
As displayed in Fig. \ref{str}, 
the Cr-O bond length of 1.87~\AA~ is about 9\% shorter than the 
Pd-O bond length.
The O-Cr-O bond angles are either 87$^\circ$ or 93$^\circ$.
The O-O-O bond angles are 60$^\circ$, 2$\times$56.9$^\circ$, and 61.6$^\circ$,
indicating nearly ideal CrO$_6$ octahedral structure.
 
Our calculations were carried out with the local spin density approximation
(LSDA) and spin-orbit coupling scheme, implemented in
two all-electron full-potential codes FPLO-9\cite{fplo} 
and Wien2k.\cite{wien2k,basis}
(Since both results are similar, we will show only ones obtained
from Wien2k unless stated otherwise.)
Additionally, the correlation effects were treated using LDA+U approach\cite{ldau}
with the double-counting scheme of the fully localized limit.\cite{fll} 
The structural parameters were optimized until forces were smaller than
1 meV/\AA.
The Brillouin zone was sampled with 20$\times$20$\times$20 $k$-mesh.

\begin{figure}[tbp]
{\resizebox{7.5cm}{7.5cm}{\includegraphics{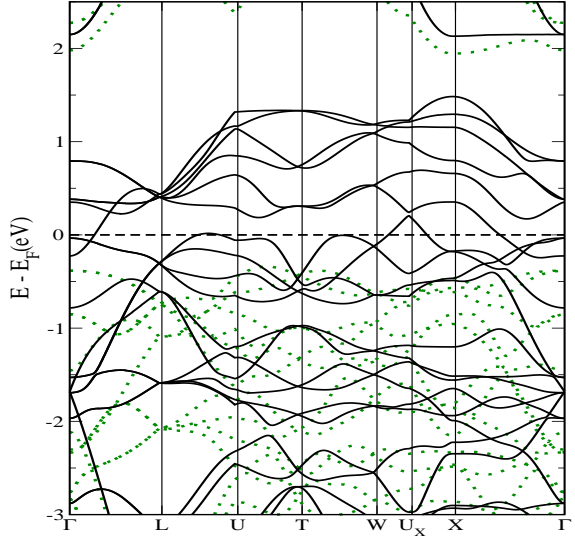}}}
\caption{ (Color online) LDA+U band structure at $U$=3 eV for Pd and 4 eV for Cr,
indicating a half-metal.
The solid and dotted lines represent the spin up and down characters,
respectively.
The symmetry points follow the notation given in Ref. \cite{LP2}. 
$E_F$ is set to zero, denoted by the horizontal dashed line.
}
\label{uband}
\end{figure}

\begin{figure}[tbp]
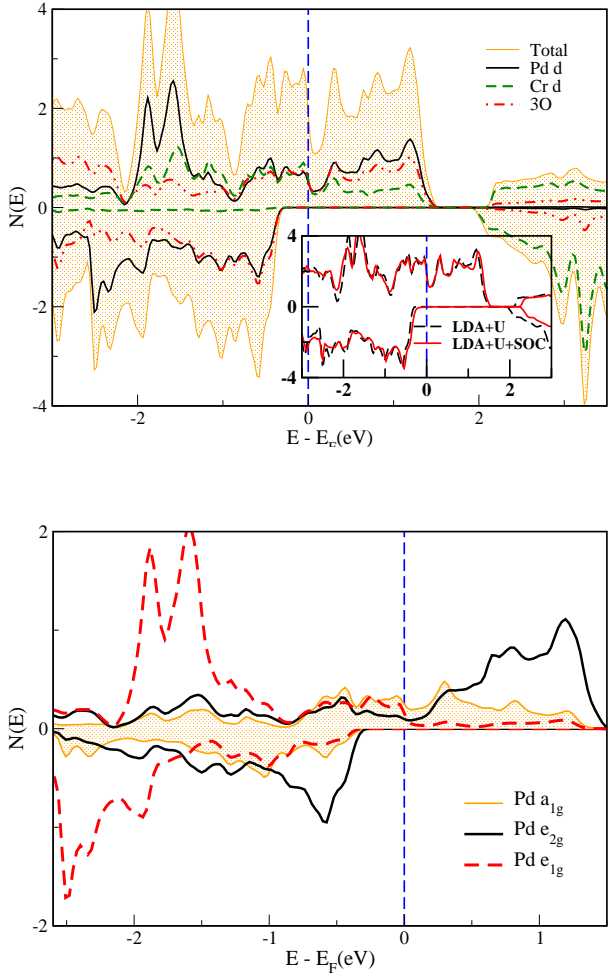

{\resizebox{8cm}{6cm}{\includegraphics{Fig3a.eps}}}
\vskip 9mm
{\resizebox{8cm}{6cm}{\includegraphics{Fig3b.eps}}}
\caption{(Color online) LDA+U densities of states (DOSs) per eV per f.u. 
 at $U$=3 eV for Pd and 4 eV for Cr.   
 Top: total and atom-projected DOSs. 
 The unoccupied manifolds lying above $\sim$2 eV are Cr $e_g$ in the spin up
 and Cr $t_{2g}$ in the spin down.
 The inset shows the compared LDA+U DOS (dashed line) with
 LDA+U+SOC DOS (solid line).
 Bottom: orbital-projected DOS of Pd $d$ states.
 The $e_{1g}$-$e_{2g}$ crystal field splittings are $\sim$3 eV 
 for the spin up and 2 eV for the down. 
 The $a_{1g}$ bands spread throughout the regime of Pd $d$ states, though
 the center is midway between the $e_{1g}$ and $e_{2g}$ manifolds.
}
\label{udos}
\end{figure}

\begin{figure}[tbp]
{\resizebox{7.5cm}{7.5cm}{\includegraphics{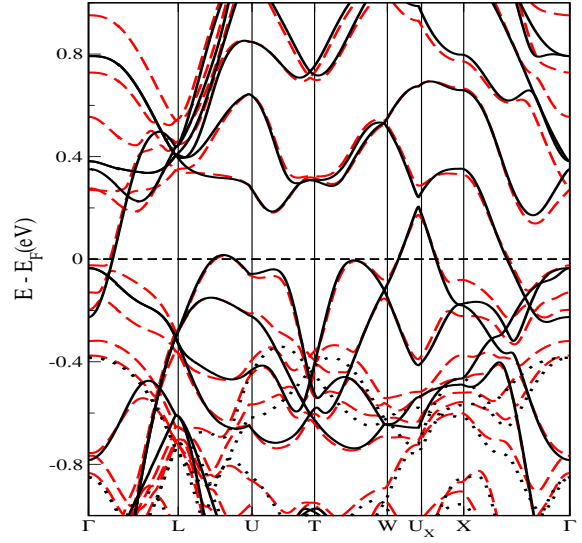}}}
\caption{(Color online) Comparison between band structures of
LDA+U (black solid and dotted lines for the spin up and down characters,
respectively) and of LDA+U+SOC (red dashed lines), near $E_F$.
In the spin down channel, the top of the valence and the bottom of 
the conduction bands shift up by 0.1 eV and 0.3 eV, respectively, 
leading to a gap of 2.5 eV.
}
\label{socband} 
\end{figure}

\begin{table}[bt]
\caption{Magnetic moments in units of $\mu_B$.
The net moment includes the moment of the interstitial regime 
not given here.
$U$=3 eV for Pd and 4 eV for Cr were used in the LDA+U and 
LDA+U+SOC calculations.
}
\begin{center}
\begin{tabular}{lccccccc}\hline\hline
     &\multicolumn{2}{c}{Cr}&~&\multicolumn{2}{c}{Pd}& 3O & net \\
                                                   \cline{2-3}\cline{5-6}
     & spin  & orbital &~ & spin & orbital & spin &    \\\hline
LSDA & 0.54 &  &~ & --0.42 &   & --0.24 &  --0.24 \\
LSDA+U & 2.08 &  &~ & --1.26 &   & --0.90 &  0.00 \\
LSDA+SOC & 0.35 & --0.01 &~ & --0.27 &  --0.05 & --0.15 &  --0.22 \\
LSDA+U+SOC & 2.13 & --0.07 &~ & --1.24 & --0.25  & --0.90 &  --0.29 \\ \hline
\end{tabular}
\end{center}
\label{table1}
\end{table}

{\it LSDA electronic structure.}
Our LSDA calculations show that the Cr moment 0.54 $\mu_B$ is antialigned 
to the Pd moment --0.42 $\mu_B$, consistent with low spin $S=\frac{1}{2}$ 
configurations in both $d^3$ Cr$^{3+}$ and $d^7$ Pd$^{3+}$ ions.
Note that these moments are considerably reduced from the formal values
due to strong hybridization and metallic character.
$N(E_F)$=4.03 states/eV-f.u. is decomposed into 15\% Cr, 35\% Pd, and 50\%
three oxygens characters (not shown here).
As shown in Table \ref{table1},
substantial oxygen moment (--0.24 $\mu_B$ for three O's) results 
in the net moment of --0.24 $\mu_B$/f.u..
This simple ferrimagnetic state has a little lower energy than 
the nonmagnetic state.

Note that the width of 3 eV of the partially occupied Pd $4d$ bands
is twice wider than the Ni $3d$ bands in NiCrO$_3$.\cite{LP2}
This suggests that PdCrO$_3$ remains metallic in ambient pressure, 
as considering even correlation effects, in contrast to NiCrO$_3$
(see below).

{\it Effects of correlation.}
Now we will address results as considering correlation effects.
We applied the on-site Coulomb repulsion $U$ to both transition metals.
The proper values of $U$ are unclear in this system, yet.
In our calculations, values of $U$ were varied in the range of $U$=2--7 eV
for both ions, while the Hund's exchange parameter $J$= 1 eV 
remained fixed.
Regardless of strength of $U$ in this range, inclusion of $U$ 
leads to a transition of the low spin to high spin configurations 
on both ions, and coincidentally a net moment exactly compensated (see below).
Besides, from comparison with experimental data,
3$-$4 eV for Cr and $\sim$3 eV for Pd in oxides have been widely 
used.\cite{LP2,srrho,Upd} 
Thus, we will focus on results obtained at $U=$4 eV for Cr and 3 eV for Pd.

The corresponding band structure and DOSs 
displayed in Figs. \ref{uband} and \ref{udos} show a half-metal.
The orbital-projected DOS of Pd $4d$ is given in the bottom panel 
of Fig. \ref{udos}.
In the spin down (insulating) channel, a gap of 2.3 eV between 
the occupied Pd $d$ and the unoccupied Cr $t_{2g}$ manifolds is open.
The occupied Pd $d$ states lie in the range of --2.8 to --0.3 eV.
The spin up (conducting) channel shows more interesting features
due to metallicity.
The occupied Pd $e_{1g}$ manifold lies on the regime of --2.2 to --0.9 eV.
The occupied Cr $t_{2g}$ manifold mainly spreads from --2 eV to $E_F$. 
The mixture of Pd $a_{1g}$, O $p$, and Cr $t_{2g}$ bands
is crossover $E_F$, indicating strong $p-d$ hybridization.
In the range of --1 to 0.2 eV, Pd, Cr, and three oxygens
distribute equally.
This strong hybridization leads to deep valleys at $-2$ eV, $-0.8$ eV, 
and 0.1 eV.
The mostly unoccupied Pd $e_{2g}$ manifold with $W$=1.5 eV 
lies just above $E_F$.

The correlation effects result in the zero net moment: 2.08 for Cr, 
--1.26 for Pd, and --0.9 for three O's (in units of $\mu_B$).
Thus, the feature of $S$=$\frac{3}{2}$ Cr$^{3+}$ and (PdO$_3$)$^{3-}$ seems 
to be reasonable, considering strong hybridization and metallicity. 
This indicates a transition of the low spin $S$=$\frac{1}{2}$ 
to high spin $S$=$\frac{3}{2}$
on both Cr and Pd ions by the on-site Coulomb repulsion, 
resulting in a precisely compensated half-metal (CHM).
(These high spin configurations follow the Hund's first rule.)

{\it Effects of Spin-Orbit Coupling.}
As usual in $3d$ or $4d$ systems, our LSDA+SOC calculations 
produce tiny orbital moments of several hundredth $\mu_B$ on both Cr and Pd
ions, implying negligible effects of SOC. 
However, in Sr$_2$RhO$_4$ the electronic structure obtained from
LDA+U+SOC calculations is consistent with 
the angle resolved photoemission (ARPES) data.\cite{srrho}  
Thus, it is of interest to investigate what happens, considering
both correlation and SOC effects.
We carried out LDA+U+SOC calculations, implemented in Wien2k. 
The resulting band structure enlarged near $E_F$ is compared with that of
LDA+U in Fig. \ref{socband}.
Although the bands having Cr character show small changes,
the largest effects occur in the Pd bands along the $\Gamma$-$L$ lines,
leading to SOC constant$\sim$0.2 eV at the $\Gamma$ point,
enlarged by a factor of two over the value of LSDA+SOC.
This shows Coulomb enhanced SOC.
These considerable effects result in a large orbital moment of $-0.25\mu_B$ 
on Pd ions above $U$=2 eV.
The Coulomb enhanced SOC is due to orbital polarization enforced by Coulomb 
repulsion.\cite{znvo,mnvo,srrho}
In spite of this substantial orbital moment,
change in the spin moments is little, as given in Table \ref{table1}.
The nearly unchanged spin moment implies tiny 
mixing with states of the opposite spin.
This is also observed in the small change in DOS by SOC,
given in the inset of Fig. \ref{udos}.
So, the half-metallicity remains unchanged even though SOC effects are considerable.
Therefore, considering the total net moment of $-0.29$ $\mu_B$,
this system is a half-metal ferrimagnet (HMFI) with large orbital moment.


{\it Discussion and Summary.}
An isovalent, but presumed cubic double perovskite La$_2$PdCrO$_6$ 
was also investigated. 
Considering results of PdCrO$_3$ and octahedral structures of PdO$_6$
and CrO$_6$, one may expect formally $S=\frac{1}{2}$ Pd$^{3+}$ and 
$S=\frac{3}{2}$ Cr$^{3+}$ ions in this system.
Thus, either HMFI or HM ferromagnet (HMFM) may be possible in this system.
Within LSDA, the spin moment of Cr is parallel to that of Pd, and
this system is a simple FM metal with some band overlap. 
Applying $U$ to both Pd and Cr ions, in the same range as in PdCrO$_3$, 
the majority $e_{g}$ manifold of Pd having $W$=2 eV is half-filled, 
resulting in HMFM instead of HMFI.
The total moment is 4 $\mu_B$, decomposed of 2.69 for Cr and 0.67 for Pd.
Contrary to PdCrO$_3$, only little orbital moments are observed
in LDA+U+SOC calculations.
This may be due to the difference in crystal field splittings between both systems,
leading to different orbital polarization.
Our calculations show that this HMFM state is stable for small distortion 
in structure, as observed in the other double perovskites.\cite{wep1,LP1}

In summary, we investigated the as-yet-unsynthesized PdCrO$_3$, isostructural
and isovalent to the trigonally distorted pervoskite-derived NiCrO$_3$, using
{\it ab initio} calculations considering both Coulomb correlations 
and spin-orbit coupling (LDA+U+SOC).
Inclusion of $U$ leads to a precisely compensated
half-metal, but unusually large orbital moment of --0.25 $\mu_B$ on Pd ion 
driven by Coulomb enhanced SOC destroys CHM.
Nevertheless, the half-metallicity still survives, 
so that this system becomes HMFI.
Our findings suggest that HM is possible even for the system 
with substantial SOC and large orbital moments.

We acknowledge W. E. Pickett for fruitful communications 
and V. Pardo for technical discussion on WIEN2k.
This research was supported by the Basic Science Research Program through
NRF of Korea under Grant No. 2011-0004683.


\begin{thebibliography}{10}
\bibitem{basis} In Wien2k, the basis size was determined by R$_{mt}$K$_{max}$=7 
 and APW sphere radii (2.05 for Pd, 1.87 for Cr, and 1.65 for O).
\bibitem{Upd} In the metallic Sr$_2$RhO$_4$, $U$=2 eV for Rh ions showed
 consistent results with the ARPES band structure.\cite{srrho} So,  
 in a metallic system, $U$$\sim$3 eV seems to be reasonable for Pd ions. 
 One may expect a higher value of $U$ for an insulating system. 
 However, even for $U$=7 eV, PdCrO$_3$ remains a compensated half-metal.
\end{thebibliography}
\end{document}